\documentclass[twocolumn,prb]{revtex4}
\usepackage{amsmath}
\usepackage{amsfonts}
\usepackage{amssymb}
\usepackage{graphics}
\usepackage{float}
\usepackage{graphicx}
\usepackage{epstopdf}
\usepackage[compact]{titlesec}
\usepackage{natbib}
\usepackage{threeparttable}
\usepackage{lipsum}

\begin{document}
\title{The influence of proximity induced ferromagnetism, superconductivity and Fermi-velocity on evolution of Berry phase in Bi$_{2}$Se$_{3}$ topological insulator}
\author{Parijat Sengupta}
\email{psengupta@cae.wisc.edu}
\affiliation{Dept of Material Science and Engineering, University of Wisconsin, Madison, WI 53706 
}

\begin{abstract}
Bi$_{2}$Se$_{3}$ is a well known 3D-topological insulators(TI) with a non-trivial Berry phase of $ \left(2n+1\right)\pi $ attributed to the topology of the band structure. The Berry phase shows non-topological deviations from $ \left(2n+1\right)\pi $ in presence of a perturbation that destroys time reversal symmetry and gives rise to a quantum system with massive Dirac fermions and finite band gap. Such a band gap opening is achieved on account of the exchange field of a ferromagnet or the intrinsic energy gap of a superconductor that influences the topological insulator surface states by virtue of the proximity effect. In this work Berry phase of such gapped systems with massive Dirac fermions is considered. Additionally, it is shown that the Berry phase for such a system also depends on the \textit{Fermi}-velocity of the surface states which can be tuned as a function of the TI film thickness. 
\end{abstract}
\maketitle

\section{Introduction}
\label{intro} 
Surface states manifest as Dirac cones protected by time reversal symmetry and impervious to external non-magnetic perturbations are formed on the surface of a 3D-topological insulator.~\cite{qi2011topological,zhang2009topological} Such surface states which impart topological insulator behaviour have been experimentally observed and established through angle-resolved photo-emission spectroscopy(ARPES) data.~\cite{chen2009experimental} While the Berry phase for materials with parabolic dispersion is trivially zero or $ 2n\pi $, it changes to a non-zero value in presence of a degeneracy in the spectrum of the Hamiltonian. A quantum mechanical system typified by a 3D-topological insulator has such a degeneracy and therefore while undergoing an adiabatic change by slowly varying a parameter of its Hamiltonian in a cyclic loop, picks up in addition to a dynamic phase, a non-trivial geometric phase. This acquired geometric phase known as the Berry phase has a value of $ \left(2n+1\right)\pi $. The non-trivial Berry phase associated with the surface states of a pristine topological insulator can be altered through a time reversal destroying external perturbation that introduces a non-topological component. In this work, the non-topological component is explicitly evaluated for a ferromagnetic exchange field and superconducting proximity effect induced perturbation and its dependence on film thickness is considered.

This paper is structured as follows: In Section~\ref{th} A, the general method of constructing surface states using a four-band continuum k.p Hamiltonian and a two dimensional Dirac Hamiltonian is introduced. The ferromagnetic proximity effect which serves as a perturbation is modeled as an exchange interaction and incorporated in the Hamiltonian. The superconducting proximity effect on a 3D-topological insulator is considered next and a BdG-type Hamiltonian is introduced. Section~\ref{th} C derives expressions for Berry phase in presence of a band gap. The dependence of Berry phase on magnitude of the exchange field, \textit{Fermi}-velocity, and overall band dispersion is demonstrated. The possibility of tuning the \textit{Fermi}-velocity to alter the non-topological component of Berry phase is also discussed. Section~\ref{res} collects results using the theoretical models developed in Section~\ref{th}. 

\section{Theory}
\label{th}
Surface and edge states in topological insulators are characterized by a linear dispersion and massless Dirac fermions. They further depend on dimensions and growth conditions of the structure that host them. Low-energy continuum models for 3D-topological insulators used in deriving results contained in this paper are described in this section. The Berry phase using analytic expressions for wave functions is derived next and takes in to account the band gap opening exchange field of the ferromagnet. 
\subsection{Model Hamiltonians for 3D-topological insulators}
The dispersion relations of Bi$_{2}$Te$_{3}$, Bi$_{2}$Se$_{3}$, and Sb$_{2}$Te$_{3}$ films are computed using a 4-band k.p Hamiltonian. The 4-band Hamiltonian~\cite{zhang2009topological} is constructed (Eq.~\ref{eqn1}) in terms of the four lowest low-lying states $ \vert P1_{z}^{+} \uparrow \rangle $, $ \vert P2_{z}^{-} \uparrow \rangle $, $ \vert P1_{z}^{+} \downarrow \rangle $, and $ \vert P2_{z}^{-} \downarrow \rangle $. Additional warping effects~\cite{fu2009hexagonal} that involve the $k^{3}$ term are omitted in this low-energy effective Hamiltonian.
\begin{align}
\label{eqn1}
H(k) = \epsilon(k) + \begin{pmatrix}
M(k) & A_{1}k_{z} & 0 & A_{2}k_{-} \\
A_{1}k_{z} & -M(k) & A_{2}k_{-} & 0 \\
0 & A_{2}k_{+} & M(k) & -A_{1}k_{z} \\
A_{2}k_{+} & 0 & -A_{1}k_{z} & -M(k) \\
\end{pmatrix}
\end{align}
where $ \epsilon(k) = C + D_{1}k_{z}^{2} + D_{2}k_{\perp}^{2}$, $ M(k) = M_{0} + B_{1}k_{z}^{2} + B_{2}k_{\perp}^{2}$ and $ k_{\pm} = k_{x} \pm ik_{y}$. The relevant parameters for Bi$_{2}$Se$_{3}$ and Bi$_{2}$Te$_{3}$ have been taken from Ref.~\onlinecite{liu2010model}.

Dispersion relationships for surface bands with linearly dispersing states in a topological insulator can also be modeled using a two-dimensional Dirac Hamiltonian. The two-dimensional Dirac Hamiltonian with additional modifications will be used while carrying out analytic derivations involving the Berry phase later in the paper. 
\begin{equation}
H_{surf.states} = \hbar v_{f}(\sigma_{x}k_{y} - \sigma_{y}k_{x})
\label{dss}
\end{equation}
Here $ v_{f}$ denotes \textit{Fermi}-velocity and $\sigma_{i} $ where $ {i = x,y} $ are the Pauli matrices. In presence of a ferromagnet with magnetization $\overrightarrow{m}$ pointing out of the plane along the \textit{z}-axis, an exchange field contribution $ \bigtriangleup_{pro}I\otimes \sigma_{z} $ must be added to the Hamiltonian. The exchange field $ \bigtriangleup_{pro} $ is introduced to quantitatively account for the proximity effect of a ferromagnet on a topological insulator.

\subsection{Hamiltonian for 3D-topological insulator and \textit{s}-wave superconductor heterostructure}
It is experimentally observed that intercalated copper in the van der Waals gaps between the Bi$_{2}$Se$_{3}$ layers yields Cu$_{x}$Bi$_{2}$Se$_{3}$ which exhibits superconductivity at 3.8 K for 0.12 $\leq$ x $\leq$ 0.15.~\cite{wray2010observation}  The exact nature of superconductivity in this alloy is yet to be fully established. Additionally, through the proximity effect at the interface between a superconductor (SC) and topological insulator, the superconductor's wave functions can penetrate the surface of a topological insulator and induce superconductivity. This induced superconductivity, by virtue of its intrinsic energy gap between the Fermi-level and the superconducting ground state offers a possible way to open a band-gap in a topological insulator.~\cite{fu2008superconducting} 

A Bogoliubov-de Gennes (BdG) Hamiltonian for a 3D-topological insulator and an \textit{s}-wave superconductor is used to compute the dispersion relationship for the topological insulator-superconductor heterostructure. In the composite Hamiltonian H$_{TS}$, $ \mu $ and $ \Delta $ denote the chemical potential pair potential respectively. The pair potential characterizes the strength of the attractive interaction potential and is a constant for an \textit{s}-wave superconductor.~\cite{heikkila2013physics,schrieffer1999theory} For the case of a TI, which is turned into a superconductor, the orbitals with opposite spin and momentum are paired. The two sets of orbitals in the 4-band TI Hamiltonian are therefore coupled by two pair potentials. The full TI-SC Hamiltonian \emph{H$_{TS}$} is written using the following basis set: {$ \vert P1_{z}^{+} \uparrow \rangle $, $ \vert P2_{z}^{-} \uparrow \rangle $, $ \vert P1_{z}^{+} \downarrow \rangle $, $ \vert P2_{z}^{-} \downarrow \rangle $, $ -\vert P1_{z}^{+} \uparrow \rangle $, $ -\vert P2_{z}^{-} \uparrow \rangle $, $ -\vert P1_{z}^{+} \downarrow \rangle $, and $ -\vert P2_{z}^{-} \downarrow \rangle $ }. A more complete description of this Hamiltonian is given in Ref~\onlinecite{sengupta2014proximity}. 
\begin{widetext}
\[
 H_{TS} =  \left( \begin{array}{cccccccc}
\epsilon+ M & A_{1}k_{z} & 0 & A_{2}k_{-} & 0 & 0 & \Delta_{1} & 0 \\
A_{1}k_{z} & \epsilon- M & A_{2}k_{-} & 0 & 0 & 0 & 0 & \Delta_{2} \\
0 & A_{2}k_{+} & \epsilon+ M & -A_{1}k_{z} & -\Delta_{1} & 0 & 0 & 0 \\
A_{2}k_{+} & 0 & -A_{1}k_{z} & \epsilon- M & 0 & -\Delta_{2} & 0 & 0 \\
0 & 0 & -\Delta_{1}^{*} & 0 & -\epsilon- M & A_{1}k_{z} & 0 & A_{2}k_{-} \\
0 & 0 & 0 & -\Delta_{2}^{*} & A_{1}k_{z} & -\epsilon+ M & A_{2}k_{-} & 0 \\
\Delta_{1}^{*} & 0 & 0 & 0 & 0 & A_{2}k_{+} & -\epsilon- M & -A_{1}k_{z} \\
0 & \Delta_{2}^{*} & 0 & 0 & A_{2}k_{+} & 0 & -A_{1}k_{z} & -\epsilon+ M \\
 \end{array} \right) - \mu I_{8 \times 8}
\label{bdg_full}
\]
\end{widetext}
 
\subsection{The Berry phase for a band gap split 3D-topological insulator}
The Berry phase is an additional geometric phase acquired by a wavefunction transported along a closed path on an adiabatic surface. For a more extensive and detailed discussion the reader is referred to standard monographs and literature.~\cite{shapere1989geometric,bohm2003geometric,chruscinski2004geometric} In a closed path $ C $ in a parameter space $ R $, Berry phase $ \gamma_{n}(C) $ is expressed as
\begin{equation}
\gamma_{n}(C) = i\oint_{C}\langle \Psi\left(r;R\right)\vert \nabla_{R}\vert \Psi\left(r;R\right)\rangle\,dR
\label{bphasecyc}
\end{equation}
where $ \vert \Psi\left(r;R\right)\rangle $ are the eigenfunctions of Schr{\"o}dinger equation $ H(R)\vert \Psi\left(r;R\right)\rangle = E_{n}(R)\vert \Psi\left(r;R\right)\rangle $.
To explicitly derive an expression for the Berry phase of a topological insulator, the wave functions of a two-dimensional Dirac Hamiltonian will be used. In presence of a ferromagnet layered on the top surface, a band gap is induced and Eq.~\ref{dss} has an additional exchange term via the proximity effect ($\Delta_{pro}$).
\begin{equation}
H_{xy} = \hbar v_{f}(\sigma_{x}k_{y} - \sigma_{y}k_{x}) + \Delta_{pro}\sigma_{z}
\label{fmdss}
\end{equation}
The eigen spectrum of Eq.~\ref{fmdss} is given as
\begin{equation}
E_{\eta}\left(k\right) = \sqrt{\Delta_{pro}^{2} + \left(\hbar v_{f}k\right)^{2}}
\label{eigspect}
\end{equation}
where $ \eta = \pm $1 denotes the helicity of the surface electrons. The wave functions of the Hamiltonian in Eq.~\ref{fmdss} is given as
\begin{subequations}
\begin{equation}
\Psi_{\eta} = \dfrac{1}{\sqrt{2}}\begin{pmatrix}
\lambda_{\eta}(k)exp(-i\theta) \\
\eta \lambda_{-\eta}(k)
\end{pmatrix}
\label{wfun1}
\end{equation}
where 
\begin{equation}
\lambda_{\eta}(k) = \sqrt{1 \pm \dfrac{\Delta_{pro}}{\sqrt{\Delta_{pro}^{2}+ \left( \hbar v_{f}k\right)^{2}}}}
\label{wfun2}
\end{equation}
\begin{equation}
\theta = tan^{-1}\left(\dfrac{k_{y}}{k_{x }}\right)
\label{theta} 
\end{equation}
\end{subequations}
To compute the Berry phase, the Berry connection $ A_{\eta}(k) = i\Psi_{\eta}^{*}\partial_{k}\Psi_{\eta} $ must be evaluated. Inserting the wave function from Eq.~\ref{wfun1}, the Berry connection expands to
\begin{subequations}
\begin{equation}
A_{\eta} = \dfrac{i}{2}\begin{pmatrix}
\lambda_{\eta}^{*}exp(i\theta) & \eta \lambda_{-\eta}^{*}
\end{pmatrix}
\begin{pmatrix}
\left( \partial_{k}\lambda_{\eta} - i\lambda_{\eta}\partial_{k}\theta\right)exp(-i\theta) \\
\eta \partial_{k}\lambda_{-\eta}
\end{pmatrix}
\label{bcurv}
\end{equation}
Simplifying the above expression,
\begin{eqnarray}
A_{\eta}(k) & = & \dfrac{i}{2}\left(\lambda_{\eta}^{*}\partial_{k}\lambda_{\eta} -i\vert \lambda_{\eta}\vert^{2}\partial_{k}\theta + \lambda_{-\eta}^{*}\partial_{k}\lambda_{-\eta}\right) \notag \\
& = & \dfrac{1}{2}\vert \lambda_{\eta}(k)\vert^{2}\partial_{k}\theta
\label{bcn2}
\end{eqnarray}
The final expression has been condensed by noting that $ \partial_{k}\lambda_{-\eta} $ evaluates exactly as $ \partial_{k}\lambda_{\eta} $ with sign reversed, therefore taken together they are equal to zero. $ \partial_{k}\lambda_{\eta} $ is worked below. 
\begin{eqnarray}
\partial_{k}\lambda_{\eta}(k) & = & \partial_{k}\sqrt{1 \pm \dfrac{\Delta_{pro}}{\sqrt{\Delta_{pro}^{2}+ \left( \hbar v_{f}k\right)^{2}}}}  \notag \\
                       & = & \eta \dfrac{1}{2\lambda_{\eta}(k)}\partial_{k}{\sqrt{\Delta_{pro}^{2}+ \left( \hbar v_{f}k\right)^{2}}}
\end{eqnarray}
\end{subequations}
The complete Berry phase can now be obtained by integrating the Berry connection $A_{\eta}(k)$ along a contour $ C $ on surface $ S $ of the topological insulator
\begin{subequations}
\begin{equation}
\gamma_{\eta} = \oint_{C} dk \cdot A_{\eta}(k)
\label{lneta}
\end{equation}
Using the expression for $ A_{\eta} $ from Eq.~\ref{bcurv}, the two-dimensional integral expands to
\begin{equation}
\begin{split}
\gamma_{\eta} = -\int dk_{x}\dfrac{1}{2}\left(1 \pm \dfrac{\Delta_{pro}}{\sqrt{\Delta_{pro}^{2}+ \left( \hbar v_{f}k\right)^{2}}} \right)\dfrac{k_{y}}{k^{2}} \\
+ \int dk_{y}\dfrac{1}{2}\left(1 \pm \dfrac{\Delta_{pro}}{\sqrt{\Delta_{pro}^{2}+ \left( \hbar v_{f}k\right)^{2}}} \right)\dfrac{k_{x}}{k^{2}}
\end{split}
\end{equation}
\end{subequations}
where using Eq.~\ref{theta}, the angular derivatives are
\begin{subequations}
\begin{equation} 
\partial_{k_{x}}\theta(k) = -\dfrac{k_{y}}{k^{2}} 
\end{equation}
\begin{equation}
\partial_{k_{y}}\theta(k) = \dfrac{k_{x}}{k^{2}} 
\end{equation}
\label{angdrv}
\end{subequations}
The Berry phase integral, after changing to polar coordinates $\left(k_{x} = kcos\theta, k_{y} = ksin\theta \right) $ for a circular energy contour evaluates to
\begin{equation}
\begin{split}
\gamma_{\eta} = \dfrac{1}{2}\int_0^{2\pi}\left(1 \pm \dfrac{\Delta_{pro}}{\sqrt{\Delta_{pro}^{2}+ \left( \hbar v_{f}k\right)^{2}}} \right)\dfrac{ksin\theta}{k^{2}}\left(ksin\theta d\theta \right) \\
+ \dfrac{1}{2}\int_0^{2\pi}\left(1 \pm \dfrac{\Delta_{pro}}{\sqrt{\Delta_{pro}^{2}+ \left( \hbar v_{f}k\right)^{2}}} \right)\dfrac{kcos\theta}{k^{2}}\left(kcos\theta d\theta \right) 
\end{split}
\end{equation}
The assumption of a circular energy contour holds good under the approximation that the integration is carried out over a constant energy surface. The presence of higher order $ k^{3} $ terms in the Hamiltonian produces a warped energy surface.~\cite{alpichshev2010stm} In such a case, the two components of the $ k $ vector must be written as $ k_{x} = k_{x}(\theta)cos(\theta)$ and $ k_{y} = k_{y}(\theta)sin(\theta)$. Since the Hamiltonian in Eq.~\ref{fmdss} is free of higher order terms, the $\overrightarrow{k}$ has no angular dependence and the integral is straightforward to evaluate yielding
\begin{equation}
\gamma_{\eta} = \pi\left(1 \pm  \dfrac{\Delta_{pro}}{\sqrt{\Delta_{pro}^{2}+ \left( \hbar v_{f}k\right)^{2}}} \right)
\label{finalberry}
\end{equation}
As evident from Eq.~\ref{finalberry}, the Berry phase in presence of a gap-opening perturbation has an additional non-topological contribution of $ \dfrac{\Delta_{pro}}{\sqrt{\Delta_{pro}^{2}+ \left( \hbar v_{f}k\right)^{2}}} $ to the topological phase of $ \pi $. The non-topological part obviously depends on the strength of exchange interaction and the \textit{Fermi}-velocity of the surface states. The \textit{Fermi}-velocity is a tunable quantity and is shown in Section ~\ref{res} that a dependence on geometry can be obtained which in turn can alter the total Berry phase.

The line integral in Eq.~\ref{lneta} can be recast by Stoke's theorem as
\begin{equation}
\gamma_{\eta} = \oint_{C} dk \cdot A_{\eta}(k) = \int_{S} dS \cdot \left(\nabla \times A_{\eta}\right) =  \int_{S} dS \cdot \overrightarrow{B}
\end{equation}
which effectively means that $ A_{\eta} $ is a fictitious vector potential. The corresponding fictitious magnetic field $\overrightarrow{B}_{fic}$ for the Berry phase evaluated in Eq.~\ref{finalberry} must have a \textit{z}-component only since all vectors are defined on a two-dimensional surface (the \textit{xy}-plane) of the topological insulator.

Using Eq.~\ref{bcn2} and expanding the curl operator in Cartesian coordinates, the fictitious magnetic field $\overrightarrow{B}_{fic}$ also known as the Berry curvature is
\begin{align}
B_{fic}(k)& =  \partial_{x}A_{\eta_{y}} - \partial_{y}A_{\eta_{x}}  \notag \\
& =  \dfrac{1}{2}\left[\left(\partial_{k_{x}}\vert \lambda_{\eta}(k)\vert^{2}\partial_{k_{y}}\theta\right)- \left(\partial_{k_{y}}\vert \lambda_{\eta}(k)\vert^{2}\partial_{k_{x}}\theta\right)\right] \notag \\
& =  \mp \dfrac{1}{2}\dfrac{\hbar^{2} v_{f}^{2}\Delta_{pro}}{\left(\sqrt{\Delta_{pro}^{2}+ \left( \hbar v_{f}k\right)^{2}}\right)^{3}}\left(k_{x}\partial_{k_{y}}\theta - k_{y}\partial_{k_{x}}\theta \right) \notag \\
& =  \mp \dfrac{1}{2}\dfrac{\hbar^{2} v_{f}^{2}\Delta_{pro}}{\left(\sqrt{\Delta_{pro}^{2}+ \left( \hbar v_{f}k\right)^{2}}\right)^{3}}
\label{bcurv1}
\end{align}
The Berry phase can thus be interpreted as the flux of a magnetic field $\overrightarrow{B}_{fic}$ across the surface $ S $ of contour $ C $ and ceases to exist when the mass inducing term $ \Delta_{pro} $ is absent.

In writing Eq.~\ref{bcurv1}, the expression $ \partial_{k_{i}}\vert \lambda_{\eta}(k)\vert^{2} $ using Eq.~\ref{angdrv} is simplified as
\begin{equation} 
\partial_{k_{i}}\vert \lambda_{\eta}(k)\vert^{2} = \mp \dfrac{2\hbar^{2} v_{f}^{2}\Delta_{pro}}{\left(\sqrt{\Delta_{pro}^{2}+ \left( \hbar v_{f}k\right)^{2}}\right)^{3}}k_{i} 
\end{equation}
where $ i = {x,y} $

\section{Results}
\label{res}
\subsection{Surface states of Bi$_{2}$Se$_{3}$ TI film}
\label{bgsplit}
\begin{figure}[h]
\includegraphics[scale=1]{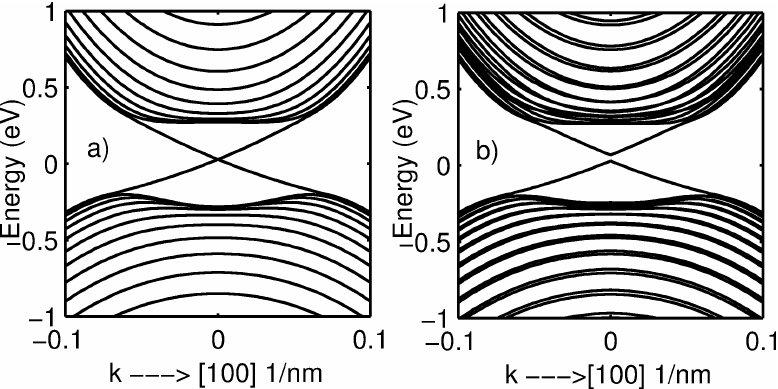}
\caption{Dispersion of a ten quintuple layer thick Bi$_{2}$Se$_{3}$ TI slab(a). Massless Dirac fermions are produced at the $ \Gamma $ point. The presence of an exchange field is equivalent to a mass term and produces massive Dirac fermions. The band gap opening in the Bi$_{2}$Se$_{3}$(b) film is roughly twice the exchange field $\left(20 \ meV \right)$.}
\label{fig1}
\end{figure}

3D topological insulators with surface states are modeled as films of finite thickness along the (111) direction. The dispersion for a ten quintuple layer thick Bi$_{2}$Se$_{3}$ $\left(E_{g\Gamma}= 0.32 \ eV\right)$ film is shown in Fig.~\ref{fig1}a. Two degenerate Dirac cones are formed at energies equal to 0.029 $\mathrm{eV}$ confirming that it is indeed a mid-gap state. The addition of an exchange field $\left(\Delta_{exc} = 20 \ meV\right) $ of a ferromagnet with an out-of-plane component along the normal(chosen to lie along the (111)-axis) splits the Dirac cones and opens a band gap. A gap approximately equal to twice the exchange energy appears and the bands acquire a parabolic character. The band gap splitting was calculated using the four-band k.p Hamiltonian (Eq.~\ref{eqn1}) in conjunction with the exchange interaction term. The magnetic proximity effect on the surface states of a topological insulator is experimentally realized by a Bi$_{2}$Se$_{3}$/EuS interface.~\cite{wei2013exchange}

\begin{figure}[t]
\includegraphics[scale=1]{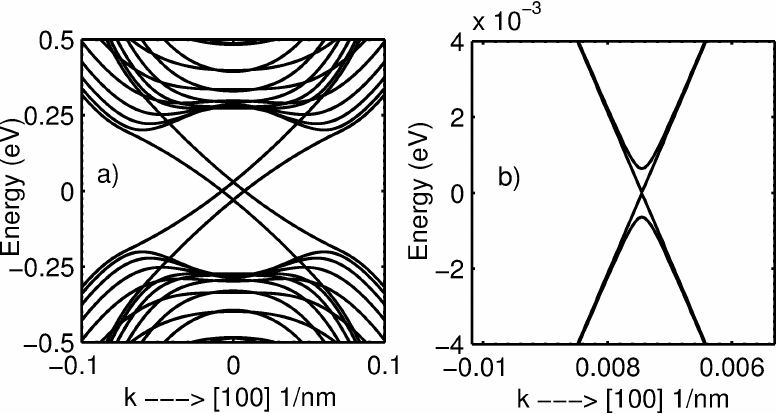}
\caption{The surface dispersion of a 40.0 $ \mathrm{nm} $ Bi$_{2}$Se$_{3}$ film when coated with an \textit{s}-wave superconductor. Fig.~\ref{fig2}a shows the overall band dispersion while Fig.~\ref{fig2}b displays the energy dispersion around the Dirac cones. The box in Fig.~\ref{fig2}a depicts an enlarged version of  the Dirac cone split because of the superconducting proximity effect. The surface with no superconductor penetration has a TI surface state.}
\label{fig2}
\end{figure}

The surface band dispersion of a 40.0 $ \mathrm{nm} $ Bi$_{2}$Se$_{3}$ film with \textit{s}-wave superconducting properties assumed to extend up to 20.0 $ \mathrm{nm} $ is shown in Fig.~\ref{fig2}. The remaining half of the film is pristine Bi$_{2}$Se$_{3}$ and possesses regular 3D-TI properties.  The assumption that superconducting behaviour is applied only to top-surface states is justified since proximity induced interactions are short-ranged effects with limited spatial penetration. The order parameters $\Delta_{1}$ and $\Delta_{2 }$ in the composite Hamiltonian $ H_{TS} $ are set to 0.34 $ \mathrm{meV} $. Since the superconductor extends only until half of the structure, the second surface still shows a Dirac cone while the top surface has an open band gap as plotted in Fig.~\ref{fig2}b. 
\subsection{The non-topological component of Berry phase}
\begin{figure}[b]
\includegraphics[scale=1]{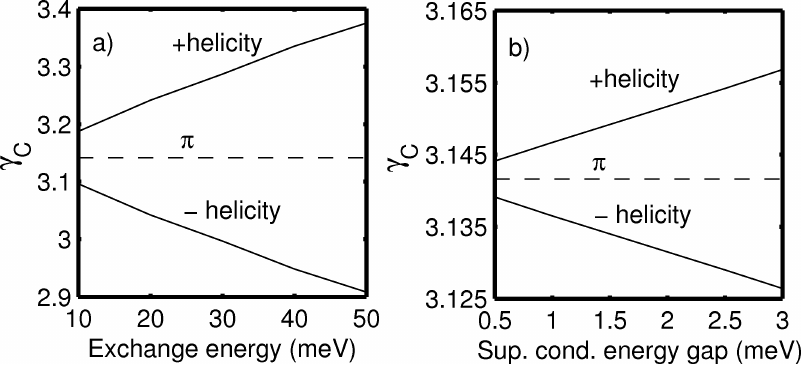}
\caption{The accumulated Berry phase for a band gap split topological insulator. The shift in Berry phase relative to the purely topological value of $\pi$ is more pronounced for a larger band gap opening as shown for the case of a ferromagnet(a). Energy gaps for \textit{s}-wave superconductors are typically smaller than the exchange energy, consequently a smaller shift in the Berry phase(b) is observed. The two values of Berry phase shown here are for electrons of opposite($ \pm $) helicity.}
\label{fig3}
\end{figure}

The destruction of zero-gapped topological insulator surface states by including a mass term has a corresponding change to the Berry phase of $ \pi $. The additional contribution here  attributable to the proximity effect of the ferromagnet or a superconductor (Eq.~\ref{finalberry}) is a function of the band gap splitting and \textit{Fermi}-velocity of the surface states. The change in Berry phase on account of inclusion of such additional proximity effects is shown in Fig.~\ref{fig3}. A constant circular energy contour of radius $\vert k \vert $ = 0.1 $ \mathrm{nm^{-1}}$ is selected as the closed surface to evaluate the Berry phase given in Eq.~\ref{lneta}. It must be noted that since a low-energy Hamiltonian has been chosen for these calculations, the energy contour must lie reasonably close to the Dirac point. The Berry phase for an energy contour with a large \textit{k}-radius must be evaluated with wave functions from a Hamiltonian with higher-order $ k^{3}$ terms. The band gap splitting results were obtained from Section ~\ref{bgsplit}. Figure.~\ref{fig3} shows that Berry phase is higher for the case of ferromagnet induced band gap compared to that of a superconductor. 

The Berry curvature or the fictitious magnetic field $\overrightarrow{B}_{fic}$ for a large exchange energy of 100 $ \mathrm{meV} $, using Eq.~\ref{bcurv1} is equal to 2.305 $\times  10^{-20}$ Tesla. The \textit{Fermi}-velocity and radius of the energy contour were set to $v_{f}$ = 6.04 $\times$ 10$^{5}$ $ \mathrm{m/s} $ and $\vert k \vert $ = 0.1 $ \mathrm{nm^{-1}}$ respectively. This is obviously a very tiny magnetic field but it is worthwhile to examine the Berry phase when the quantum system with Dirac fermions is placed in a large external magnetic field $ \overrightarrow{B}_{ext} $. The exchange energy $\Delta_{pro} $ in Eq.~\ref{finalberry} is now replaced by the Zeeman splitting $ g\mu_{B}m_{j}B_{ext} $. If $ g\mu_{B}m_{j}B_{ext} \gg \hbar v_{f}k $, the Berry phase equals $ 2n \pi $, $ n \in Z $. The spin of a Dirac fermion rotates in the \textit{xy}-plane due to spin-momentum locking but under a strong magnetic field, the spin aligns with the field direction and the Berry phase changes to $ 2n \pi $. 

\begin{figure}[t]
\includegraphics[scale=0.85]{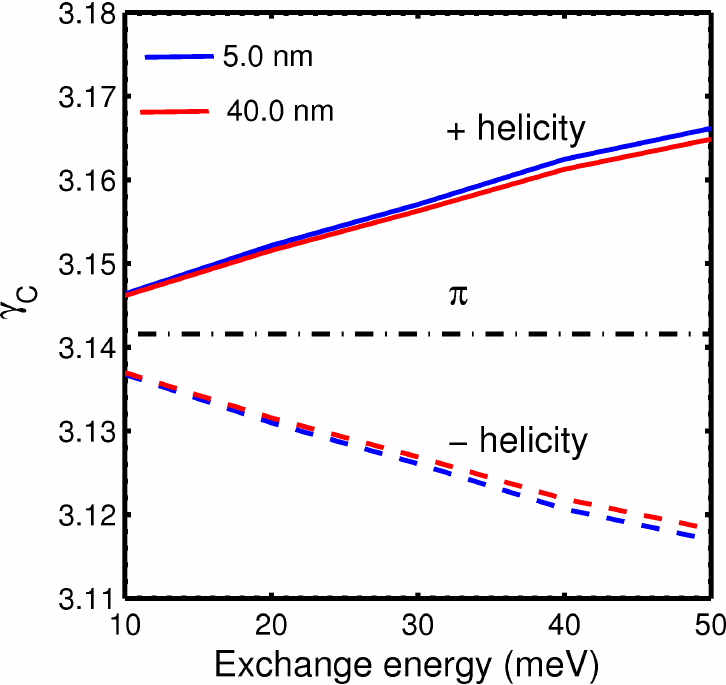}
\caption{The complete Berry phase in a mass-gapped system depends on the \textit{Fermi}-velocity of surface electrons. The increase in Berry phase is greater for a 5.0 $ \mathrm{nm} $ compared to a thicker 40.0 $ \mathrm{nm} $ film. The band gap is introduced through an exchange interaction term arising on account of the proximity effect of a ferromagnet.}
\label{fig4}
\end{figure}

Finally, as evident from Eq.~\ref{finalberry}, the \textit{Fermi}-velocity impacts the overall the change in Berry phase. For two Bi$_{2}$Se$_{3}$ slabs of thickness 5.0 $ \mathrm{nm} $ and 40.0 $ \mathrm{nm} $, the \textit{Fermi}-velocity is computed to be 5.671 $\times$ 10$^{5}$ $ \mathrm{m/s} $ and 6.04 $\times$ 10$^{5}$ $ \mathrm{m/s} $ respectively. The \textit{Fermi}-velocity was determined using the standard result $ v_{f} = \dfrac{1}{\hbar}\dfrac{\partial E}{\partial k}$ and the derivative evaluated numerically from the dispersion plot obtained by diagonalizing the four-band Hamiltonian (Eq.~\ref{eqn1}). These values are close to experimentally determined \textit{Fermi}-velocity.~\cite{qu2010quantum} The non-topological component of the Berry phase corresponding to these \textit{Fermi}-velocities is compared in Fig.~\ref{fig4}. The shift in Berry phase is slightly more for the 5.0 $ \mathrm{nm} $ film with a lower \textit{Fermi}-velocity compared to the thicker 40.0 $ \mathrm{nm} $ film. Apart from thickness of the film, an external electric field can be used to tune the velocities of the surface electrons and obtain a different Berry phase. This alternative method hasn't been pursued in this work though.
\section{Conclusion}
The zero-gap surface-state electrons of a topological insulator are massless Dirac fermions which pick up a geometric phase of $\pi$ when they complete a closed-loop path. This non-trivial phase of $ \pi $ can be altered if the massless Dirac fermions acquire mass and surface state bands are gapped. The massive Dirac fermions contribute a non-topological component which is a function of the band gap opening induced on account of the proximity effect of a ferromagnet or an \textit{s}-wave superconductor. The calculation of Berry phase is not just an esoteric idea in condensed matter physics but finds wide application in areas as diverse as macroscopic electric polarization in ferroelectric materials~\cite{resta1994macroscopic,king1993theory} to the well-known Jahn-Teller effect.\cite{grosso2014solid,spaldin2012beginner} Dirac fermions acquire a finite anomalous velocity in presence of a finite Berry curvature giving rise to anomalous quantum Hall effect.~\cite{xiao2010berry}. In this work, an isotropic Hamiltonian for the surface states is chosen without explicitly including the higher-order terms that lead to well-known warping effects. It is expected that the Berry phase would increase with warping but such calculations are intended for later work. 
\begin{acknowledgements}
We thank late Prof. Gabrielle.F. Giuliani from the Dept. of Physics at Purdue University for introducing one of us (PS) to Berry phase and its myriad manifestations in condensed matter physics. We also thank Intel Corp. for support during early stages of this work. 
\end{acknowledgements} 

\bibliographystyle{apsrev}

\end{document}